\documentstyle[aps,epsf]{revtex}  

\newcommand{\D}{ /\! \! \! \! D}
\begin{document}        
\draft
\preprint{UM-TH-99-03 }       

\baselineskip 14pt
\title{A non-local OPE for hard QCD processes near the elastic limit}
\author{Ratindranath Akhoury and Michael G. Sotiropoulos\footnote{Talk presented at the APS Division of Particles and Fields conference, UCLA, January 1999} }
\address{Randall Laboratory of Physics, University of Michigan, Ann Arbor, MI
48109}
\author{George Sterman}
\address{Institute for Theoretical Physics, SUNY at Stony Brook, Stony Brook,
NY 11794}
\maketitle              

\begin{abstract}        
A leading twist expansion in terms of bilocal operators is proposed for the
structure functions of deeply inelastic scattering near the elastic limit $x
\rightarrow 1$, which is also applicable to a range of other hard
quasi-elastic processes. Operators of increasing dimensions contribute to
logarithmically enhanced terms,
 which are suppressed by corresponding powers of
$1-x$. For the longitudinal structure function
in moment $(N)$ space
 all the
logarithmic contributions of order $\ln^k N/N$ are shown to be resummable in
terms of the anomalous dimension of the leading operator in the expansion.
\
\end{abstract}   	

\section{Introduction}               

The treatment of hard processes in QCD perturbation theory involves the
factorization of the short from the long distance dynamical regimes,
characterized 
by
mass scales $Q$ and $\Lambda_{\rm QCD}$ respectively.
Such processes become quasi-elastic in those regions of phase space where
a 
final state invariant mass $M$ becomes much smaller than $Q$.
For the perturbative treatment of these three-scale processes the condition
$M^2 \gg \Lambda^2_{\rm QCD}$ is also necessary. Typical quasi-elastic
processes are deeply inelastic scattering (DIS) at large Bjorken $x$,
Drell-Yan production near the partonic threshold and thrust in $e^+$-$e^-$
annihilation near the two-jet limit.
We argue that
existing 
methods for studying quasielastic processes can be
extended in a systematic
way, 
 via a non-local operator expansion to be
introduced below.
The DIS longitudinal structure function will be considered for definiteness.

The longitudinal structure function $(F_L)$, like any DIS observable at large
$Q$,
may be treated by 
the Wilson OPE,
organized through the light cone expansion or, equivalently,
factorization formulas.  
Given regulator $(\epsilon)$ and factorization scheme $(\overline{\rm MS})$,
the leading twist factorization at scale $\mu^2$ can be written as
\begin{equation}
F_L(x, Q^2, \epsilon) = \bar C_L(x, Q^2/\mu^2, \alpha_s(\mu^2)) \otimes f(x,
\alpha_s(\mu^2),  \epsilon)+ {\cal O}(\Lambda^2_{\rm QCD}/Q^2) \, ,
\label{flOPEfac}
\end{equation}
where we define the convolution in longitudinal momentum as
\begin{equation}
h(x) \otimes f(x) =  \int_0^1 d x_1 \; \int _0^1 dx_2 \; \delta(x-x_1 x_2) \;
h(x_1) \; f(x_2) \, ,
\label{conv}
\end{equation}
and the perturbative expansion of
any function $R$ 
as
$R = \sum_{n=0}^\infty (\alpha_s/\pi)^n R^{(n)}$.
The coefficient functions can be computed in perturbation theory and the
results to the second order in the coupling are \cite{DDKS,SG,ZvN}
\begin{eqnarray}
&\ &\bar C_L^{(0)}(x, Q^2/\mu^2) = 0 \, , \hspace{2cm}
\bar C_L^{(1)}(x,Q^2/\mu^2) = C_F x,
\nonumber \\
&\ & \bar{C}_L^{(2)}(x,Q^2/\mu^2) =  \left[
\frac{1}{2} \left(P_{qq}^{(1)} \otimes \bar{C}_L^{(1)}\right)(x)
-\frac{1}{4}\beta_1 \,
\bar{C}_L^{(1)}(x) \right]
\ln \frac{Q^2}{\mu^2} +\bar{C}_L^{(2)}(x,1) \, ,
\label{coeffs}
\end{eqnarray}
with $P_{qq}$ the quark splitting function and $\beta_1$ the one-loop
coefficient of the beta function.
The coefficient function at $Q^2 = \mu^2$ is
\begin{equation}
\bar{C}_L^{(2)}(x,1) =
\frac{1}{2} C_F^2\ln^2(1-x) +
\left[\frac{9}{4} -2 \zeta(2)\right]C_F^2 \ln (1-x)
+ \left[ (\zeta(2)-1)C_F C_A -\frac{1}{4} \beta_1 C_F \right] \ln(1-x)
+{\rm reg} \, .
\label{CL2}
\end{equation}
Only
terms non-analytic at $x=1$  
are displayed in the above
equation.
Our objective is to show that these terms can be captured by expectations of
appropriately defined operators, 
 and that they can be resummed to all orders in
perturbation theory.
It is instructive at this point to compare the above coefficient function with
the corresponding one for $F_2$ to ${\cal O}(\alpha_s)$,  
\begin{equation}
\bar C_2^{(1)}(x, Q^2/\mu^2) = \frac{1}{2} P_{qq}^{(1)} \ln\frac{Q^2}{\mu^2} +
4 C_F \left( \frac{\ln (1-x)}{1-x}\right)_+
- 3 \frac{1}{(1-x)_+} + 4 C_F \ln(1-x) + {\rm reg}.
\label{C21}
\end{equation}
We note the absence of plus distributions in the case of $F_L$. The
longitudinal structure function has only logarithmic divergences to all orders
in perturbation theory. Upon taking the Mellin transform
\begin{equation}
\tilde F(N, Q^2) = \int_0^1 dx \, x^{N-1} \,  F(x, Q^2) \, ,
\label{Mellin}
\end{equation}
the leading logarithms of $F_L$ are of order $\ln^k N/N$ in moment space.
Near the elastic limit $(N \rightarrow \infty)$ they are power suppressed
relative to the order $\ln^k N$ terms coming from the plus distributions of
$F_2$.
For this reason,   
only recently
has 
the Sudakov factorization
formalism
been 
extended to capture purely logarithmic threshold corrections
\cite{ASS1}.

\section{The non-local OPE in the elastic limit}

The Sudakov factorization formalism of perturbative QCD applies to hard
processes in the quasielastic regime $Q^2 \gg M^2 \gg \Lambda_{\rm QCD}^2$ and
resums terms of order $\ln^k N$, where $N$ is the moment with respect to the
variable $(Q^2-M^2)/Q^2$.
We shall refer to this formalism as
the leading-jet approximation,
and we briefly
summarize it here before extending it to the case of $F_L$.
(For a recent review and references see Ref.~\cite{CoLaSte}).

Every DIS observable is obtained as a particular projection of the hadronic
tensor
\begin{equation}
W_{\mu \nu}(p, q) =  \frac{1}{4 \pi}
\int d^4 y \; {\rm e}^{-iq\cdot y} \;
\langle p | j_\mu^\dagger(0) \, j_\nu(y) |p \rangle \, .
\label{wtens}
\end{equation}
Structure functions are obtained as $F_L = P_L^{\mu \nu} W_{\mu \nu}$,
$F_2 = P_2^{\mu \nu} W_{\mu \nu}$ with the projectors
\begin{equation}
P_L^{\mu \nu} = \frac{8 x^2}{Q^2} p^\mu p^\nu \, ,
\ \ \ \ \
P_2^{\mu \nu} = -\eta^{\mu\nu} +\frac{3}{2} P^{\mu\nu}_L \, .
\label{projectors}
\end{equation}
The
leading-jet
expansion
of $F_2$ leads to the following factorization formula
\begin{equation}
F_2(x, Q^2) =  |H_2(Q^2)|^2 \,
\int_x^1 dx' \int_0^{x'-x} dw\; J\left((x'-x-w)Q^2\right)\; V(w) \; \phi(x')
\; \Big( 1+{\cal O}(1-x) \Big) \, .
\label{f2fac}
\end{equation}
All three factors in the longitudinal momentum convolution integral are
defined as expectation values of non-local operators. Explicitly, the soft
function is
\begin{equation}
V(w) = \int dy^- {\rm e}^{-iw y^- p \cdot \bar v} \;
\langle 0| \Phi^\dagger_v(0, -\infty) \;
\Phi_{\bar{v}}(0, y^- \bar v) \;
\Phi_{v}(y^- \bar v, -\infty) |0 \rangle \, ,
\label{Vdef}
\end{equation}
the initial state jet function is
\begin{equation}
\phi(x') =
\int d y^- \, e^{-i x^\prime y^- p\cdot \bar v} \langle p| \psi(0)
\not{\bar{v}} \;
\Phi_{\bar{v}}(0,y^-) \, \bar{\psi}(y)|p\rangle \otimes V^{-1}\, ,
\label{phidef}
\end{equation}
and the final state jet function is
\begin{equation}
J\left((1-z)Q^2\right)=
\int d^4 y \, e^{-i(q+z p)\cdot y}
\langle 0| \Phi^\dagger_v(0, -\infty) \; \psi(0) \,
\bar{\psi}(y) \;  \Phi_{v}(y, -\infty) |0 \rangle \otimes V^{-1} \, .
\label{jdef}
\end{equation}
We have defined the $v^\mu = \delta^\mu_+$ light-cone vector in the direction
of motion of the incoming parton and $\bar v^\mu$ as the opposite $(-)$
direction. We have also introduced the Wilson line operator
\begin{equation}
\Phi_v(x+tv, -\infty) =
P \exp\left[ -i g_s \int_{-\infty}^t d s \,
v^\mu \, A_\mu( x+s v ) \right] \, .
\label{Phidef}
\end{equation}
The perturbative expansion of the Wilson line operator generates the Feynman
rules for the eikonal interaction between a jet of fast moving partons in the
$v^\mu$ direction and the soft gluons.
The presence of the $V^{-1}$ factor in the jet definitions removes any purely
soft contributions from these functions.
In a perturbative calculation of the jets, this corresponds to the prescription
of applying soft subtractions in the contributing diagrams order by order.

In this operator language we can see that the factorization formula
Eq.~(\ref{f2fac}) comes from a rearrangement of the fields that define the
currents to which the DIS probe couples (in our case the electromagnetic
current $j^\mu = \bar \psi \gamma^\mu \psi$).
Indeed, by
``wiring"  
 one of the fermion fields to the initial state and the other
to the final state we obtain
\begin{equation}
P_r^{\mu \nu} W_{\mu \nu} = P_r^{\mu \nu} {\rm FT}^{(4)}_q
\langle p | \bar \psi(0) \gamma_\mu \psi(0) \bar \psi(y) \gamma_\nu \psi(y)
|p\rangle
\approx {\rm FT}^{(1)}_{xp} \langle p | \bar \psi(0) \psi(y^-) | p \rangle
\otimes {\rm FT}^{(4)}_{q+x p} P_r^{\mu \nu} \gamma_\mu \langle 0| \psi(0)
\bar \psi(y) | 0 \rangle \gamma_\nu \, ,
\label{wired}
\end{equation}
where we have introduced ${\rm FT}^{(1)}_{xp}$ and ${\rm FT}^{(4)}_q$ as
shorthands for the one- and four-dimensional Fourier transforms. In this,
leading jet, 
approximation, the $F_r, \, r=2,L$, 
 DIS structure functions are written as
convolutions 
 of two jets, which are
identified as the initial and final
state jets defined in Eqs~(\ref{phidef},\ref{jdef}). Note that this procedure
would give a vanishing result for $F_L$, 
 because the projection of the final
state matrix element with $P_L$, Eq.~(\ref{projectors}), is zero by the Dirac
equation of motion for the on-shell incoming quark.
$F_L$ vanishes in the leading-jet approximation, and  
the coefficient functions of $F_L$ are
less singular 
in the
small parameter $1-x$ relative to the ones of $F_2$, as we noted in the
introduction.
The operator formalism indicates how it can be
expanded to capture such terms that are suppressed by powers of $1-x$. To this
end we introduce higher dimensional operators with the quantum numbers of the
electromagnetic current and rearrange their fields between initial and final
states. Let $O^\mu_i(y) = \bar \psi(y) \Xi^\mu_i(y)$ be an operator of
dimension higher than 3. Then, as in Eq.~(\ref{wired}), we obtain
a new contribution 
\begin{eqnarray}
P_r^{\mu \nu} W_{\mu \nu} &=& P_r^{\mu \nu}
\; |H_L(Q^2)|^2\; 
{\rm FT}^{(4)}_q
\langle p | \bar \psi(0) \Xi_{\mu ,i}(0) \, \Xi^\dagger_{\nu,i}(y) \psi(y)
|p\rangle \nonumber\\
&\approx&
|H_L(Q^2)|^2\; 
{\rm FT}^{(1)}_{xp} \langle p | \bar \psi(0) \psi(y^-) | p \rangle
\otimes {\rm FT}^{(4)}_{q+x p} P_r^{\mu \nu}
\langle 0| \Xi_{\mu,i}(0) \Xi^\dagger_{\nu,i}(y) | 0 \rangle  \, .
\label{wiredXi}
\end{eqnarray}
with $H_L(Q^2)$ a new coefficient function. 
The second factor in the convolution product above is a {\it new} jet
function. At each mass dimension
for the operator $O^\mu_i$ we generate
a class of jet functions whose contributions are suppressed by one factor of
$1-x$ relative to
those of 
lower dimension level. Therefore, 
 an expansion in powers of $1-x$ reduces to dimensional counting,
subject to 
the equations of motion of the incoming
and outgoing partons. This is the non-local (bilocal) operator product
expansion near
the elastic limit \cite{ASS1}.
In the language of effective field theories, we have ``integrated out"
the dynamics between the scales $(1-x)Q^2$ and $Q^2$, which
generates in a new operators $O_i$ in  an effective Hamiltonian
for the still-perturbative dynamics at the intermediate scale.

The bilocal OPE in the elastic limit is a method of capturing corrections of
order $(1-x)^n$ in the final state interactions. This expansion is still
leading twist in $Q^2$. This
may be 
understood by observing that, 
 if the dimension
of the effective operator $O^\mu_i$ is $3+n$, then a field of dimension
$3/2+n$ is emitted
into 
the final state and the phase space integral scales as
$(1-x)^n Q^{2n}$. It is therefore the phase space integral that absorbs the
factor $1/Q^{2n}$
associated with the coefficient functions for
the higher dimensional operators, leaving the
factor $(1-x)^n$, which generates an expansion in $1-x$.

Let us apply the above procedure to $F_L$. We seek to define a new jet
function that captures the leading power contributions in $1-x$. The dimension
4 operators with the quantum numbers of the current are
\begin{equation}
O_{2a}^\mu = \bar v^\mu \bar \psi \D_\perp \psi \, ,
\ \ \ \
O_{2b}^\mu = v^\mu \bar \psi \D_\perp \psi \, ,
\ \ \ \
O_{2c}^\mu = \bar \psi D_\nu \sigma_\perp^{\mu \nu} \psi \, ,
\ \ \ \
O_{2d}^\mu = \bar \psi D^\mu \psi \, .
\label{theops}
\end{equation}
It is easily seen that $O_{2b}$ and $O_{2c}$ have vanishing projection when
contracted with $P_L^{\mu \nu}$. They do not contribute to $F_L$ but can
contribute to $F_2$ at the subleading level. $O_{2d}$ has
a 
longitudinal
projection that is proportional to $\bar \psi (v \cdot
\stackrel{\leftarrow}{D}) \psi$. After integration by parts this term vanishes
by the equation of motion of $\psi$ at $x=1$. We therefore conclude that the
only operator of dimension 4 that will contribute to $F_L$ at the leading
level is $O_{2a}$, with longitudinal projection
\begin{equation}
v_\mu O^\mu_{2a} =\bar \psi \stackrel{\leftarrow}{\D}_\perp \psi \,.
\label{2aproj}
\end{equation}
This results in the jet function for $F_L$
\begin{equation}
J^\prime((1-z) Q^2) =
 \left(\frac{1}{4\pi} \; \frac{8 x^2}{Q^2} \right) \;
 {\rm FT}^{(4)}_{q+zp} \;
\langle 0| \Phi_{v}^\dagger (0, -\infty) \; \D_\perp \psi(0) \;
\bar{\psi}(y) \stackrel{\leftarrow}{\D}_\perp \; \Phi_v (y, -\infty) |0
\rangle \otimes V^{-1} \, ,
\label{jprimedef}
\end{equation}
where we
have 
restored the Wilson lines for manifest gauge invariance and
have 
divided
out any purely soft contributions. The resulting factorization formula for
$F_L$ is similar to the one for $F_2$.
\begin{equation}
F_L(x, Q^2) =  |H_2(Q^2)|^2 \,
\int_x^1 dx' \int_0^{x'-x} dw\; J^\prime\left((x'-x-w)Q^2\right)\; V(w) \;
\phi(x') \; \Big( 1+{\cal O}(1-x) \Big) \, .
\label{fLfac}
\end{equation}
In the $F_L$ resummation formula the anomalous dimension of $J^\prime$ will
appear.

\section{Beyond the eikonal approximation - resummation}

The factorization formula for $F_L$ at large Bjorken $x$ was derived via the
non-local OPE near the elastic limit in the previous section. It is however
instructive
also to 
study the problem from the point of view of perturbation
theory diagrams. This approach
yields the Feynman rules for computing the
anomalous dimensions of the higher jet functions.

The starting point of the diagrammatic approach is the observation that both
$F_2$ and $F_L$ are different projections of the {\it same} hadronic tensor
$W^{\mu \nu}$. Any non-analytic terms in their expansion in massless
perturbation theory arise from well known configurations in loop momentum
space, the Landau pinch singular surfaces of $W^{\mu \nu}$. We can therefore
use the infrared power counting and the jet-soft factorization techniques on
which general QCD factorization theorems are based (for a review see
Ref.~\cite{CoSoSte}). Given the light-cone direction vectors $v^\mu$ and $\bar
v^\mu$, normalized as $v \cdot \bar v =1$, the momenta of the incoming quark
and the outgoing jet $p$ and $\bar p$ respectively are
\begin{equation}
p^\mu = \frac{Q}{\sqrt{2}} v^\mu \, ,
\ \ \ \
\bar p^\mu = (1-x) \frac{Q}{\sqrt{2}} v^\mu + \frac{Q}{\sqrt{2}x} \bar v^\mu
\, .
\label{moms}
\end{equation}


\begin{figure}[ht]	
\centerline{\epsfxsize 2.0 truein \epsfbox{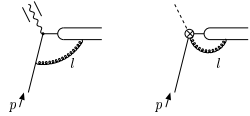}}
\caption[]{
\label{Fig1}
\small The factorization of the $\bar v$-collinear gluon $(l)$ from the hard
subgraph generates the effective vertex $\otimes$. Parallel lines around the
photon denote contraction with $p^\mu$.}
\end{figure}

Consider the configuration shown in Fig.~1, in which the depicted gluon
belongs to the $\bar v$-jet.
The fermion line carrying momentum $p-l$
is far off-shell and belongs to the hard scattering subdiagram  $H_L$.
If, however, we were to factorize the
top 
collinear gluon
by eikonalizing the fermion propagator,  
 then this, in
general leading, IR contribution would
vanish  
in $F_L$ because of
the contraction with $p^\mu$
from $P^{\mu\nu}$ 
at the current vertex. Note also in the case
where the gluon is soft, the leading eikonal contribution is similarly
projected out. Therefore, the expansion will start with terms that are
subleading in the usual eikonal approximation. Expanding the diagram in the
gluon's virtuality we obtain
\begin{equation}
g_s \frac{1}{-2 p\cdot l +l^2} \not p (\not p -\not l) \gamma^\lambda
\approx g_s \left( \gamma_\perp^\lambda - l_\perp \cdot \gamma_\perp
\frac{v^\lambda}{v \cdot l} \right)
\equiv O^\lambda(l) \, ,
\label{vertex}
\end{equation}
where we have dropped terms vanishing by the incoming quark's equation of
motion. An effective vertex to ${\cal O}(g_s)$ has emerged. This vertex,
unlike the usual eikonal ones, probes the transverse momentum of the struck
quark. It satisfies the following algebaic properties, that are useful in
diagrammatic calculations:
\begin{equation}
O^\lambda(l) l_\lambda =0, \ \ \
O^\lambda(l) v _\lambda = 0, \ \ \
O^\lambda(l) \bar v_\lambda = - g_s \not l_\perp \frac{1}{v \cdot l} \, ,
\ \ \
O^\lambda(l) O_\lambda(l^\prime) = - 2 g_s^2, \ \ \
O^\lambda(l) ... O_\lambda(l^\prime) = g_s^2 \gamma_\perp^\lambda ...
\gamma_{\perp \lambda} \, .
\label{properties}
\end{equation}
Moreover, it is noted that the vertex $O^\lambda$ is the ${\cal O}(g_s)$
expansion in momentum space of the operator $\Phi_v(0, -\infty) \D_\perp
\psi(0)$, i.e. half of the bilocal operator that defines $J^\prime$,
Eq.~(\ref{jprimedef}). It can be shown by induction that the above
construction generalizes
to
any number of collinear gluons 
\cite{ASS2}.
The factorization of the final state collinear partons from the initial state
jet defines a final state jet function exactly as in Eq.~(\ref{jprimedef}).

Resummation
may be thought of
a consequence of factorization.
\cite{CoLaSte}  
 Once an observable is written
as a product of effective operators,  
 then the UV renormalization of these
operators generates the evolution of the coefficient functions of the
observable. Convolutions become simple products in moment space where the
Mellin transform of $F_L$, Eq.~(\ref{Mellin}), can be written as
\begin{eqnarray}
\tilde F_L(N, Q^2, \epsilon) &=&
\left|H_L\left(\frac{(p\cdot \bar{v})^2}{\mu^2},
\frac{(\bar{p}\cdot v)^2}{\mu^2},\alpha_s(\mu^2) \right)\right|^2
\nonumber \\
&\ & \times \frac{1}{N}
\tilde{J}^\prime \left( \frac{Q^2}{N \mu^2},
\frac{(\bar{p} \cdot v)^2}{\mu^2}, \alpha_s(\mu^2) \right) \,
\tilde{V}\left( \frac{Q^2}{N^2 \mu^2},  \alpha_s(\mu^2)\right) \,
\tilde{\phi}\left( \frac{Q^2}{N^2 \mu^2}, \frac{(p \cdot \bar{v})^2}{\mu^2},
\alpha_s(\mu^2), \epsilon \right) \, .
\label{flfacconv}
\end{eqnarray}
Here, the factor of $1/N$ has been exhibited explicitly,  
\begin{equation}
\frac{1}{N} \, \tilde{J}^\prime\left( \frac{Q^2}{N \mu^2}, \frac{(\bar{p} \cdot
v)^2}{\mu^2},
\alpha_s(\mu^2)\right)
= \int_0^1 dz \, z^{N-1} \,
J^\prime \left(\frac{(1-z) Q^2}{x \mu^2}, \frac{(\bar{p} \cdot v)^2}{\mu^2},
 \alpha_s(\mu^2) \right) \, .
\label{Mellindef}
\end{equation}
The treatment of the $J^\prime$ evolution proceeds in the usual way
\cite{CoLaSte}. The jet function satisfies two equations. The first is the
Lorentz covariance equation,
which 
describes how the jet varies
in response to 
changes
in the light-cone direction with respect to which it has been defined,
and the 
second is the jet's own renormalization group equation,
\begin{equation}
\frac{\partial}{\partial \ln (\bar{p}\cdot v)^2}
\ln \tilde{J}^\prime \left(\frac{Q^2}{N \mu^2},
\frac{(\bar{p}\cdot v)^2}{\mu^2},\alpha_s(\mu^2) \right) =
K\left(\frac{Q^2}{N \mu^2}, \alpha_s(\mu^2)\right)
+ G\left( \frac{(\bar{p} \cdot v)^2}{\mu^2}, \alpha_s(\mu^2) \right),
\label{cov}
\end{equation}
\begin{equation}
\frac{d}{d \ln \mu^2}
\ln \tilde{J}^\prime \left(\frac{Q^2}{N \mu^2},
\frac{(\bar{p}\cdot v)^2}{\mu^2}, \alpha_s(\mu^2) \right)
= -\frac{1}{2}\gamma_{J^\prime}(\alpha_s(\mu^2)).
\label{rg}
\end{equation}
The right-hand side of the covariance equation is renormalization group
invariant and the $K$ factor renormalizes as
\begin{equation}
\frac{d}{d \ln \mu^2}
K\left( \frac{Q^2}{N \mu^2}, \alpha_s(\mu^2) \right) =
-\frac{1}{2}\gamma_K(\alpha_s(\mu^2)),
\label{rgK}
\end{equation}
where $\gamma_K$ is 
the well known cusp anomalous dimension in $\overline {\rm MS}$ scheme, 
\begin{equation}
\gamma_K(\alpha_s) = \frac{\alpha_s}{\pi} C_F
+\left(\frac{\alpha_s}{\pi}\right)^2\left[ C_F C_A
\left(\frac{67}{36}-\frac{\pi^2}{12}\right) -\frac{5}{18} C_F N_f\right]
+{\cal O}(\alpha_s^3).
\label{cusp}
\end{equation}
Eqs~(\ref{cov},\ref{rg}) can be solved simultaneously subject to the initial
condition
\begin{equation}
\tilde J^\prime(1, 1, \alpha_s(Q^2)) = C_F \frac{\alpha_s(Q^2)}{\pi} +{\cal
O}(\alpha_s^2(Q^2)) \, 
,
\label{initcon}
\end{equation}
from the Mellin transform, Eq.\ (\ref{Mellindef}), of 
$J^\prime$ to lowest order ${\cal O}(\alpha_s)$.
To
this order,  
 there is only one contributing diagram, depicted in Fig.~2.

\begin{figure}[ht]	
\centerline{\epsfxsize 2.0 truein \epsfbox{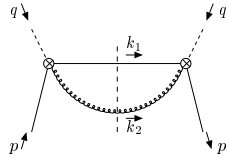}}
\caption[]{
\label{Fig2}
\small The lowest order contribution to the jet function $J^\prime$.}
\end{figure}

Substituting the solution of the jet equations
(\ref{cov}) and (\ref{rg}) 
into the factorization formula
(\ref{flfacconv}), 
 we obtain the
resummed expression 
\begin{eqnarray}
&\ &\tilde F_L(N, Q^2, \epsilon) \, = \,
C_F \frac{\alpha_s(Q^2)}{\pi N} \,
(\tilde V   \tilde \phi )(\mu^2 = Q^2/N, \epsilon)
\nonumber \\
&\ & \times
\exp \left[ \frac{1}{2} \int_{Q^2/N}^{Q^2}
\frac{d \mu^2}{\mu^2}
\left( \ln N \, \gamma_K(\alpha_s(\mu^2)) +\gamma_{J^\prime}(\alpha_s(\mu^2))
+ 2\beta(\alpha_s) \frac {\partial}{\partial \alpha_s} \ln \tilde
J^\prime(1,1,\alpha_s(\mu^2)) \right) \right]
+{\cal O}\left(\frac{\ln^0 N}{N}\right) \, .
\label{flsol}
\end{eqnarray}
Apart from
the jet $J'$, 
$N$ dependence is
also found from 
in the factor
$\tilde V \tilde \phi$
in Eq.\ (\ref{flfacconv}).  
 This dependence, 
however, cancels 
in the coefficient
function $C_L$ once we divide $\tilde F_L$ by the Mellin transform of the
parton distribution function $\tilde f$, Eq.~(\ref{flOPEfac}). In
Eq.~(\ref{flsol}), the contributions coming from the UV
renormalization of the jet function $J^\prime$ are contained in
$\gamma_{J^\prime}$ and are kept distinct from the renormalization of the
coupling, contained in the $\beta$-function term of the exponent. This
distinction is important for the
bookkeeping  
of the renormalization terms in
the perturbative calculation of the anomalous dimensions.

In summary, we have
shown 
that the extension of the
bilocal OPE to 
the case of $F_L$ modifies the Sudakov resummation formula essentially only
through 
the introduction of the anomalous dimension of the new jet function
$J^\prime$. The cusp anomalous dimension $\gamma_K$ contains all universal
soft radiation effects.

\section{The jet anomalous dimension to ${\cal O}(alpha_s )$}

Inspecting the resummation formula, Eq.~(\ref{flsol}), we see that the only
new ingredient is the anomalous dimension $\gamma_{J^\prime}$.
In this section, 
we outline its calculation to ${\cal O}(\alpha_s)$. Because $F_L$ itself and
$J^\prime$ start at ${\cal O}(\alpha_s)$, $F_L^{(1)} = J^{\prime(1)} = \bar
C_L^{(1)} = C_F x$, we need to compute $J^\prime$ to ${\cal O}(\alpha_s^2)$.
Using the rules for the effective operator $O^\lambda$ of the previous section
we identify the second order diagrams shown in Fig.~3.


\begin{figure}[ht]	
\centerline{\epsfxsize 3.0 truein \epsfbox{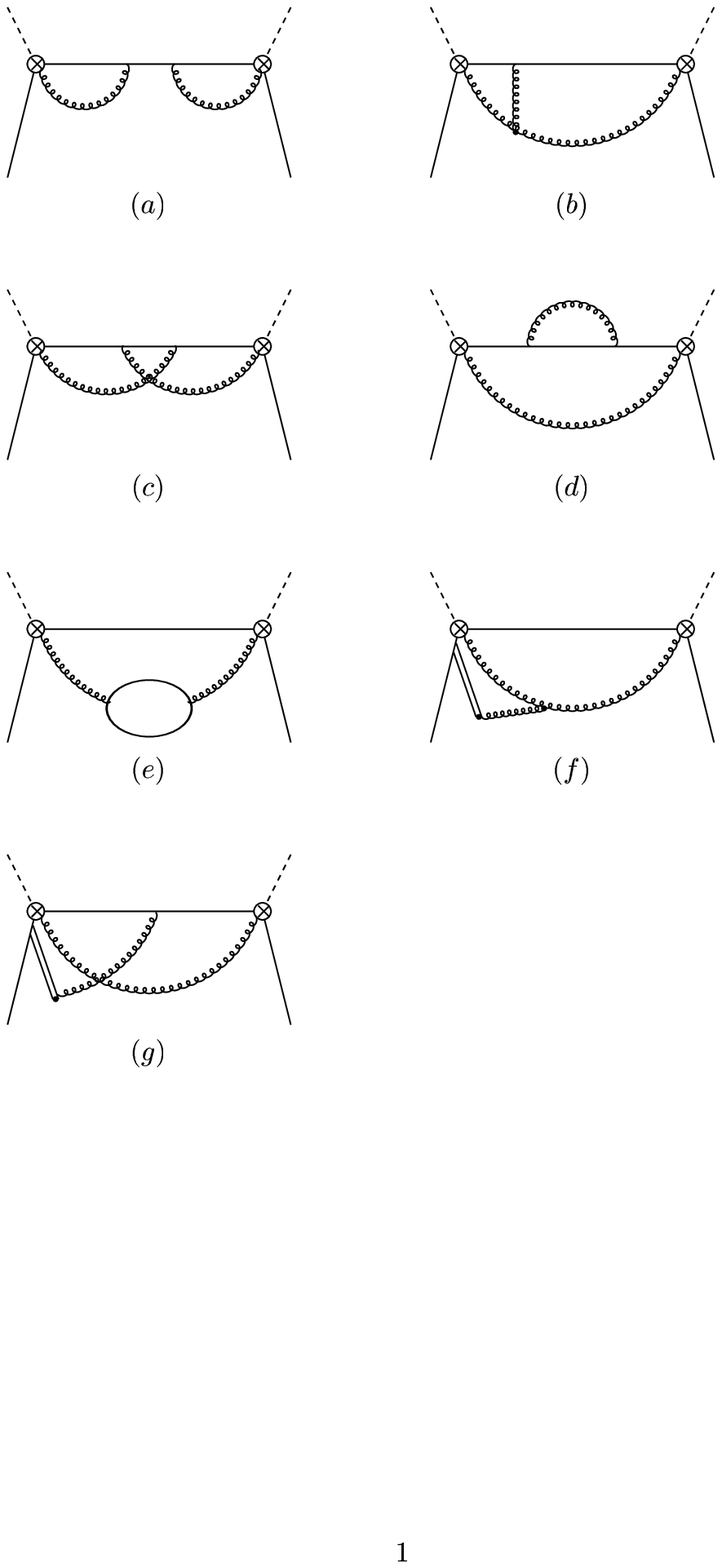}}
\caption[]{
\label{Fig3}
\small The ${\cal O}(\alpha_s^2)$ diagrams for the jet function $J^\prime$.}
\end{figure}

First we compute both the UV and IR singular parts of the diagrams. Then we
perform an IR subtraction as dictated by Eq.~(\ref{jprimedef}). The infrared
part to be subtracted out is $J^{\prime (1)} \otimes V^{(1)}$ and we note that
the soft function $V$ is the same for
$F_L$ and $F_2$. 
After the
subtraction,  
 only UV simple poles survive and from these we construct as usual
the anomalous dimension $\gamma_{J^\prime}$. The result is
\begin{equation}
\gamma_{J^\prime}(\alpha_s) = \frac{\alpha_s}{\pi}
\left[ \frac{9}{2}C_F -2 C_A - 4 \zeta(2) \left(C_F-\frac{C_A}{2}\right)
\right] +{\cal O}(\alpha_s^2).
\label{gammaJtotno}
\end{equation}
Comparison of the resummation with the fixed order results listed in the
introduction is made by first writing the resummation prediction for the
coefficient function to ${\cal O}(\alpha_s^2)$
\begin{equation}
\tilde C_L^{(2)} = \frac{C_F}{2 N} \left[ \gamma_K^{(1)} \ln^2 \frac{N}{N_0}
-\left(\gamma_{J^\prime}^{(1)} - \frac{1}{2} \beta_1 \right) \ln \frac{N}{N_0}
\right]
\label{coeffpred}
\end{equation}
with $N_0 = e^{-\gamma_E}$. Upon substituting for the anomalous dimensions
$\gamma_K$ and $\gamma_{J^\prime}$ from Eqs.~(\ref{cusp}, {\ref{gammaJtotno})
we obtain the Mellin transform of the fixed order result in Eq.~(\ref{CL2}).

\section{Summary}
We have shown how the light-cone expansion for hard QCD processes near the
elastic limit can be extended into a general
bilocal 
OPE that captures power
terms 
suppressed in $1-x$.
This operator
expansion is based on purely dimensional arguments, 
 but it can also be derived
from the analysis of the perturbative diagrams in the infrared limit. The
result is a Sudakov resummation formula for the observable, 
 in which the
anomalous dimensions of the higher dimensional jet functions enter the
exponent. We
have 
analyzed the DIS longitudinal structure function at large Bjorken
$x$ as a concrete example of this operator expansion near the elastic limit.
The anomalous dimension of the $F_L$ final state jet was computed to first
order in $\alpha_s$, 
 and the resummation formula for the coefficient function
was shown to agree with the fixed order calculations.

We emphasize that our analysis is leading twist in $Q^2$
and is consistent with the usual 
light-cone expansion. It resums terms that are of order $\ln^kN/N, \, k > 0$.
This approach can be extended to the case of $F_2$, where such terms are
present, but
are 
subleading relative to order $\ln^k N, \, k \ge 0$ terms. Indeed
we expect that the operator expansion presented here is relevant for all hard
QCD processes in the quasi-elastic limit.


\begin{references}  

\bibitem{DDKS} A. Devoto, D. W. Duke, J. D. Kimel and G. A. Sowell,
Phys. Rev. {\bf D30,} 541 (1984).

\bibitem{SG} J. S\'{a}nchez Guill\'{e}n {\it et al.},
Nucl. Phys. {\bf B353,} 337 (1991).

\bibitem{ZvN}  E. B. Zijlstra and W. L. van Neerven,
Nucl. Phys. {\bf B383,} 525 (1992).

\bibitem{ASS1} R. Akhoury, M. G. Sotiropoulos and G. Sterman,
Phys. Rev. Lett. {\bf 81,} 3819 (1998).

\bibitem{CoLaSte}  H. Contopanagos, E. Laenen, and G. Sterman,
Nucl. Phys. {\bf B484,} 303 (1997).

\bibitem{CoSoSte} J.C. Collins, D.E. Soper and G. Sterman, in {\it
Perturbative Quantum Chromodynamics}, editor A.H. Mueller, World Scientific,
Singapore, 1989.

\bibitem{ASS2} R. Akhoury, M.G. Sotiropoulos and G. Sterman, in preparation.

\end{references}
\end{document}